# Open Orchestration Cloud Radio Access Network (OOCRAN) Testbed*

Marti Floriach-Pigem
Dept. of Signal Theory and Communication
Barcelona Tech-UPC
Castelldefels, Spain
mfloriach90@gmail.com

Guillem Xercavins-Torregrosa
Dept. of Signal Theory and Communication
Barcelona Tech-UPC
Castelldefels, Spain
guillemxercavins@gmail.com

Vuk Marojevic
Wireless@VT, Bradley Dept. of Electrical and Computer Engineering
Virginia Tech
Blacksburg, VA, USA
maroje@vt.edu

Antoni Gelonch-Bosch
Dept. of Signal Theory and Communication
Barcelona Tech-UPC
Castelldefels, Spain
antoni@tsc.upc.edu

## ABSTRACT

The[1] Cloud radio access network (C-RAN) offers a revolutionary approach to cellular network deployment, management and evolution. Advances in software-defined radio (SDR) and networking technology, moreover, enable delivering software-defined everything through the Cloud. Resources will be pooled and dynamically allocated leveraging abstraction, virtualization, and consolidation techniques; processes will be automated using common application programming interfaces; and network functions and services will be programmatically provided through an orchestrator. OOCRAN, oocran.dynu.com, is a software framework that is based on the NFV MANO architecture proposed by ETSI. It provides an orchestration layer for the entire wireless infrastructure, including hardware, software, spectrum, fronthaul and backhaul. OOCRAN extends existing NFV management frameworks by incorporating the radio communications layers and their management dependencies. The wireless infrastructure provider can then dynamically provision virtualized wireless networks to wireless service providers. The testbed's physical infrastructure is built around a computing cluster that executes open-source SDR libraries and connects to SDR-based remote radio heads. We demonstrate the operation of OOCRAN and discuss the temporal implications of dynamic LTE small cell network deployments.

## 1 INTRODUCTION

We are witnessing an explosive expansion of Cloud computing and the use of data centers for providing diverse types of commercial and non-commercial services. The benefits come from the increased freedom of development, scalability and personalization of systems and services, and the rapid deployment of new functionalities, products and services. Virtualization is the enabler of resource sharing and its use reaches far beyond processing, storage and wired networking [1]. Research and development (R&D) is now incorporating virtualization technology into wireless communications networks with a special emphasis on the physical layer using software-defined radio (SDR) technology [2] [3].

Wireless communications and Cloud computing have important differences that need to be considered when merging the two. The Cloud provides computing, storage, data, applications, and other services that are hosted on some remote physical resources, such as a data center or a server connected to the Internet. Wireless communications networks have strict Quality of Service (QoS) requirements and exhibit a high degree of heterogeneity of equipment, system configurations, and services. Moreover, currently deployed 4G long-term evolution (LTE) networks fall short in terms of capacity and latency for enabling the tactile Internet, autonomous control of terrestrial and aerial vehicles, and massive connectivity of the Internet of Things. In addition, the end-to-end wireless network infrastructure and resources that deliver the services are often proprietary and heavily regulated. Despite the technological and non-technological barriers that exist today, using and improving

Cloud computing technology is needed to enable the evolution of wireless communications and networking technology and services [4].

The business model behind the Cloud radio access network (C-RAN) is based on the Infrastructure-as-a-Service (IaaS) model. A pool of physical resources, which include antenna sites, networking components and processors, is virtualized and offered to mobile virtual network operators (MVNOs) to build their networks as most suitable for the services they intend to offer [5]. As technology evolves, MVNOs will be able to load software images of wireless networks that can be dynamically customized and adapted to the changing operational conditions. That is, a MVNO can request more resources or release resources on the fly to "reimage" its network. Whereas this seems like a logical extension of Cloud computing, several research challenges remain.

Telecommunications infrastructure providers have been showing growing interest in incorporating Cloud computing technology in their service networks. The feasibility and benefits of using data centers and the Cloud led to gradually moving their network infrastructure to a virtual environment. During the 2013 Mobile World Congress several institutions agreed on standardizing the administration of the virtualized network infrastructure.

An important push in this direction has come from the European Telecommunications Standards Institute (ETSI). ETSI clarifies the scope of network functions virtualization (NFV) and defines standard specifications that are meant to fulfill the operational and management requirements of next generation wireless networks [6]. NFV decouples the physical network equipment from the network functions. A NFV implementation is understood as an instance, or virtual network function (VNF), and is completely software-defined. Reference [7] demonstrates VNFs by executing the 3G and 4G core networks (CNs) in virtual machines (VMs). OpenStack and the kernel-based virtual machine (KVM) hypervisor provide the overall virtualization and management layers. Reference [8] presents the design, implementation, and evaluation of two LTE CN architectures, one being based on the principles of software-defined networking (SDN) and the other on NFV. Reference [9] describes the virtualization process of a base station and CN, whereas [10] performs a comparison between the proposed SDN and NFV solutions in mobile radio environments.

Fig. 1 illustrates the ETSI NFV management and orchestration (MANO) architecture, which is composed of three main building blocks:

- The **Orchestrator** manages the overall network and is responsible for including new services and VNF packages.
- The **VNF Manager** oversees the lifecycle management of VNFs on the NFV infrastructure (NFVI) according to the specifications provided by the orchestrator.
- The **Virtualized Infrastructure Manager (VIM)** manages the compute, storage and network resources of the NFVI.

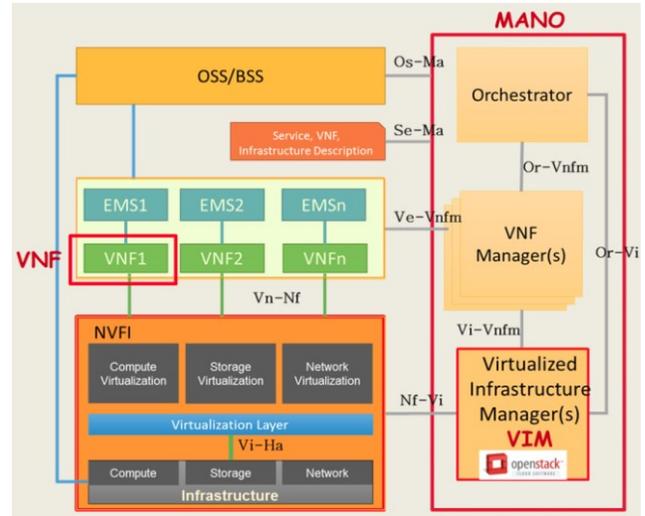

**Figure 1: ETSI NFV MANO architecture.**

During the past ten years a plethora of Cloud computing solutions have been developed for providing IaaS. These tools allow managing servers as well as network infrastructure. Among the most popular IaaS frameworks are OpenStack [11], Eucalyptus [12], and the Ubuntu Cloud infrastructure [13]. OpenStack has become a prominent solution because of its capability to satisfy the network infrastructure providers' needs in terms of massive computing and storage and complex networking. It can be considered as a VIM in the ETSI MANO reference architecture.

In our research we assume that wireless infrastructure customers, typically MVNOs, lease resources as needed or periodically from an IaaS provider to build their networks [5]. These virtualized wireless networks will be tailored to the services they provide by using virtual wireless network components as their building blocks, rather than dealing directly with the computing, spectrum, remote radio heads (RRHs), antennas, fronthaul/backhaul, and other physical resources. The MVNO thus deploys a network of virtual resources with agreed communications, computing and networking capabilities.

We have identified a gap between experiments in isolated radio environments and production-ready virtualized wireless networks. In order to leverage research and education in this field and facilitate transition to practice, we introduce the Open Orchestration C-RAN (OOCRAN) testbed, oocran.dynu.com. OOCRAN extends existing NFV management frameworks by incorporating the radio communications layers and some of their management dependencies. It provides an orchestration layer for the entire wireless infrastructure—hardware and software—so that the wireless infrastructure provider (WIP) can dynamically provision virtualized wireless networks to wireless service providers (WSPs) and satisfy the instantaneous communications needs. This paper presents our C-RAN testbed and illustrates the key features and example applications of OOCRAN. Section 2 summarizes the testbed highlights. Section 3 presents the OOCRAN architecture and hardware components, whereas

Section 4 introduces the software layers. Section 5 discusses some use cases for deploying SDR LTE networks on the testbed to enable research and education. We conclude the paper with an outlook on research on virtualized wireless networks, adaptive network services, and ultra-dense 5G networks.

## 2  TESTBED OBJECTIVES AND HIGHLIGHTS

The testbed objectives were defined to support experimental research and education on virtualized wireless networks in a laboratory and campus environment. These are:
- Provide a cloud computing platform for executing VNFs,
- Enable real-time resource management across heterogeneous resource pools,
- Provide high flexibility, capacity, and extensibility, and
- Be compliant with the ETSI NFV MANO architecture.

The testbed design should (1) leverage the virtualization capability of C-RANs, as opposed to solely centralizing the baseband processing, and (2) enable the generation and real-time processing of RF signals in a Cloud computing environment.

OOCRAN uses the IaaS reference model as the basis of its design. It fills the orchestration and management gap of next generation virtualized wireless networks by providing a clear and easy way to dynamically configure and assemble the building blocks of virtualized wireless networks and deploy them on shared infrastructure. It achieves this by virtualizing physical resources and managing shared access to limited computing and radio resources under a given policy. The modular design and isolation between hardware and software facilitates testbed upgrades. The OOCRAN framework manages the virtual-to-physical resource mapping and allows creating new wireless networks as supported by the available physical resources [14].

OOCRAN is an orchestration layer that follows the ETSI MANO architecture for creating, coordinating and managing wireless networks. An OOCRAN user can take the role of a WIP, a WSP or a wireless test provider (WTP). This enables analyzing different ways of splitting the resource deployment, management and maintenance responsibilities and evaluating resource access policies using a single framework. In the role of a WIP, OOCRAN can create complete communications systems by chaining several VNFs and delivering them to the WSP. Acting as a WSP, OOCRAN can precisely manage the wireless infrastructure and introduce proper management policies to optimize the system behavior for the given environmental conditions and service demands. Acting as a WTP, OOCRAN facilitates creating specific operating conditions and test procedures and uses Cloud-based system monitoring tools to analyze the system. In other words, a WTP creates local or remote wireless laboratories that are tailored to the R&D needs.

The OOCRAN framework has been developed using OpenStack Newton release as the VIM with the Neutron and Heat options. The Neutron extension allows OOCRAN users to customize remote access to the virtual networks they create. Heat, on the other hand, allows saving a defined infrastructure as a description file, facilitating quick deployment or deletion.

The KVM hypervisor provides an open source high performance platform.

The core of the OOCRAN management environment is based on the Django framework, which facilitates the creation of complex, database-driven Web sites and thus simplifies the development of new use cases. The user interface is based on Web services, which allow activating schedulers, queues, alarms, monitoring and scripting tools to properly manage VNFs and their configurations. These tools enable building a detailed control layer of the virtual infrastructure and applying various operational policies that take into account the state of resources and services.

The wireless system is implemented combining several VMs, each performing specific VNFs. These VMs execute SDR waveforms and run on general-purpose processors with access to SDR hardware and RF equipment. They are capable of generating and processing real-time signals and interfacing radio links and CN functionalities. The currently available system is LTE, created as a fork of the srsLTE software library [15]. All the source code is open and released under the AGPL license. It can be freely downloaded from the OOCRAN project repository [16].

The OOCRAN testbed is capable of running in simulation mode as well as creating an emulated or real wireless network. A combination of simulated and real system components allows rapid prototyping and testing in controlled RF environments.

## 3  ARCHITECTURE AND EQUIPMENT

Fig. 2 depicts the OOCRAN testbed architecture. It features several components that enable implementing, managing, and analyzing virtualized wireless networks. A computing cluster emulates the data center of the C-RAN. The RRHs are accessed through the radio aggregation unit (RAU), using the terminology of the Next Generation Fronthaul Interface (NGFI) for 5G [17]. Together they form the remote radio system. RF instruments can be attached for signal or spectrum analysis, among others.

Fig. 3a shows a photo of the testbed hardware. The roles and capabilities of the hardware components are discussed in continuation.

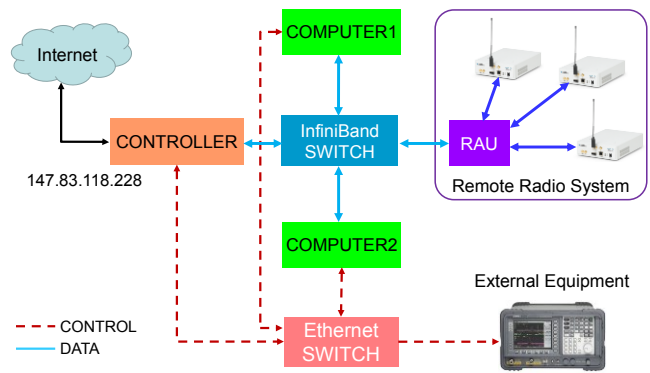

**Figure 2: OOCRAN testbed architecture.**

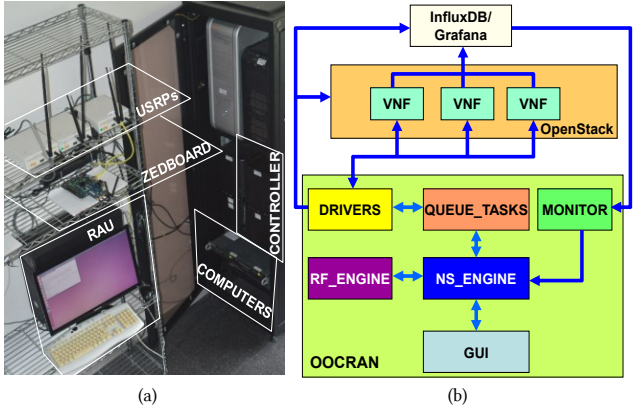

Figure 3: OOCRAN testbed hardware (a) and software (b) modules.

## 3.1 Computing Cluster

The testbed is built around a computing cluster. Using the OpenStack terminology, one PC acts as the Controller and the other two as Computers, all running Ubuntu 16.04. The Controller features a 3rd generation Intel i7 8-core processor running at 2.5 GHz and using 8 GB of RAM. Its mission is to administrate the C-RAN. Among other things it manages the VM lifecycle, user-defined networks, and virtual routers. The two other Computers are two rackmount workstations, model SuperServer 6018TR-TF from SuperMicro. Each has two 2.6 GHz Intel Xeon 12-core processors, two Gigabit Ethernet ports, and one 56 Gbps InfiniBand port. These workstations host the VMs. They carry out the heavy computation, processing and forwarding the incoming data flows to the RF and CN components.

The InfiniBand Switch IS5022 is an 8-port non-blocking and unmanaged 40 Gbps switching system that is capable of delivering 640 Gbps bandwidth with 100 ns port-to-port latency. It acts as a general switch of the testbed's high speed network that connects the computing cluster with the remote radio system.

## 3.2 Remote Radio System

One additional PC—Intel i7-6700 8-core processor operating at 3.4 GHz with 32 GB of RAM—is included in the testbed as the RAU. It connects to the data center via the InfiniBand switch and to the RRHs through Gigabit Ethernet. The RAU handles the necessary data forwarding between the two networks.

Five RRHs are currently available as part of the OOCRAN testbed. We use the N210 model of Universal Software Radio Peripherals (USRPs) from Ettus Research. These USRPs allow sampling at up to 50 Mega-samples per second (MS/s), are capable of generating or capturing RF signals below 6 GHz and connect to the processing center through Gigabit Ethernet. The remote radio system also features the following two RRHs: the ZedBoard with the AD-FMCOMMS3-EBZ daughterboard, capable of capturing 56 MHz of instantaneous RF bandwidth, and LimeSDR, which extends the bandwidth to 61.44 MHz.

## 3.3 External Equipment

RF instruments, such as spectrum analyzers, as well as additional computing or radio equipment can be connected to the testbed through the Ethernet switch or the RAU. This allows extending the experimental and RF analysis capabilities of the testbed.

## 4 SOFTWARE LAYERS

The OOCRAN software layers have been designed to provide a wireless management framework that extends the functionalities of ETSI MANO. These layers facilitate the focus on designing optimized management algorithms, called *actuators*. Our software framework, depicted in Fig. 3b, consists of six functional modules: GUI, MONITOR, NS_ENGINE, RF_ENGINE, QUEUE_TASKS and DRIVERS.

### 4.1 GUI

The graphical user interface (GUI) provides a user-friendly operating environment to facilitate the interaction between the user (human network operator) and the virtual infrastructure or network services (NSs). The GUI has been developed using Python 2.7 and Django. Django uses a model-view-controller, which facilitates making modification to the provided code. This allows saving the NSs and VNFs and defining actuators that can execute one or several tasks such as perform a partial reconfiguration or modify the lifecycle of the VNFs according to the state of the NS.

### 4.2 MONITOR

The MONITOR module configures the third-party programs Grafana/InfluxdB. InfluxDB is a framework that captures the state of the virtualized wireless infrastructure (VWI) and saves it in a database. OOCRAN accesses this database, processes the saved states by searching for predefined patterns, and starts the VWI reconfiguration process if the state matches the specified conditions. Grafana is a plotting tool that is used to plot the desired data. By using both frameworks we are capable of capturing data from VNFs, processing the data and creating customized graphs and alarms. This allows processing and exposing the state of a VNF (computational load, active users, waveform type, etc.).

MONITOR captures the alarms that the Grafana/InfluxdB programs generate from the state of the NSs and VNFs. All alarms include a unique identifier related to a specific actuator. The module checks the credentials and the alarm identifier and, when both align, sends a command to NS_ENGINE to execute the corresponding actuator. This could, for example, trigger a partial reconfiguration of a set of VNFs.

### 4.3 NS_ENGINE

The NS_ENGINE module manages the NSs and the actuators. It decides about the NS/VNF lifecycle or the actions following certain conditions (time, alarm, input, etc.). This module is supported by RabbitMQ, which manages process queues and, among others, allows executing tasks asynchronously. When the

NS_ENGINE needs to apply a change to the infrastructure, it updates its own database and sends a new task to the queue manager (QUEUE_TASKS). When the new task arrives, it is executed by the DRIVERS module that uses third-party application programming interfaces (APIs) to perform the update.

### 4.4 RF_ENGINE

The RF_ENGINE module manages the pool of radio resources (RF channels, transmission power levels, transmitter masks, etc.). It creates slices of spectrum and assigns them to different VNFs to avoid RF interference among coexisting radios and networks. NS_ENGINE interacts with RF_ENGINE when creating a new NS or VWI.

These modules are compatible with the VIMs of OpenStack or Vagrant; this allows building and maintaining portable virtual software development platforms.

### 4.5 Interfaces

The OOCRAN software base is a fork of the Django framework and, therefore, all Django APIs can be used. OOCRAN incorporates APIs from OpenStack, Vagrant, InfluxDB and Grafana to create, delete, reconfigure and monitor the NSs. Fig. 3b shows the interfaces among the OOCRAN modules and other management components. The information exchanges between modules are done through calls to classes and their methods. Third-party program drivers use HTTP RESTful APIs, which allows installing RabbitMQ, InfluxDB, Grafana, and OpenStack on different computers. Alarms use Webhook, an HTTP callback that detects changes in the working conditions of a program.

## 5 EXAMPLE SCENARIOS

The assessment of management frameworks or the design of resource management strategies can be done using simulated scenarios. On the other hand, the virtualization of physical resources and the associated management issues, such as resource slicing, isolation, and dynamic provisioning, require real RF equipment. VMs can be effectively deployed to host LTE base stations (eNodeBs) and user equipment (UEs). Additional VMs can simulate RF channels or connect to physical RRHs. This allows switching between simulated and real wireless links or using mixed links. Here we show the deployment of an LTE VWI and discuss VWI reconfiguration.

### 5.1 Virtualized LTE System

The scenario for creating the minimum infrastructure for the LTE downlink signaling is shown in Fig. 4.

VNFs are attached to two different subnets with different functionalities: a) the DataFlow Network carries the data instances and b) the Management Network carries the control instances. The IP address assignment is done by means of DHCP in both cases. The DataFlow Network interconnects VNFs with external RRHs for sending or receiving IQ samples. The Management Network provides connectivity between OOCRAN/InfluxDB/SDN and the VNFs to send information about operational states. More precisely, the state of the eNodeB VNFs is provided and VNF reconfiguration triggered using a secure shell (SSH) connection. The VNF application metrics are sent to the InfluxDB database using the Python API. Alarms from InfluxdB/Grafana are displayed at the following URL: http://oocran:8000/alerts/messages.

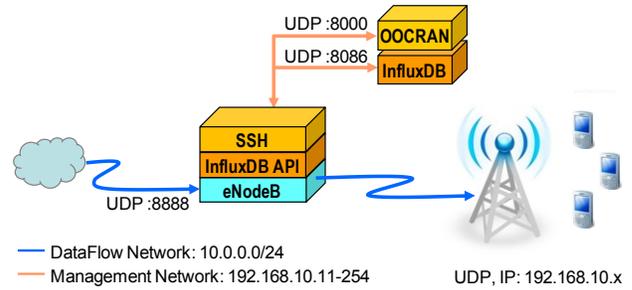

**Figure 4: Real LTE RF link setup.**

OOCRAN assembles the required infrastructure building blocks for any real or simulated communications network and configures the corresponding VNF interfaces.

The signal generated by the eNodeB transmitter instance is sent to the RRH. The radiated RF signal is captured by another RRH and sent to a spectrum analyzer instance. The spectrum analyzer instance executes the UHD_FFT tool from GNU Radio (www.gnuradio.org). Fig. 5 captures the Horizon front end from OpenStack, the Ubuntu instance screen and the UHD_FFT tool showing a capture of the LTE downlink spectrum generated by the eNodeB instance. This spectrum exhibits enough quality, that is, enough signal strength with respect to the noise floor.

The UE receiver, a UE instance with its USRP and processing unit, can demodulate the signal with a resulting block error rate below 0.1, which indicates proper system operation.

### 5.2 Dynamic VWI Deployment

This scenario describes the dynamic deployment of a VWI, more precisely, a small cell-based wireless infrastructure (30 m cell radius). Here the OOCRAN management layer takes the role of a WIP. It designs and generates a suitable VWI on demand (using slices of physical resources) to satisfy the WSP needs. The virtual wireless network maps to physical resources and includes slices of RRHs, spectrum, transmission and processing power, among others.

Some general assumptions of this scenario are
- Use of omnidirectional antennas,
- Line of sight links and free space path loss model,
- RRHs can operate at various frequencies,
- eNodeB sends data to multiple subscribers at the same time and assigns different bit rates,
- 1.4 MHz LTE channel bandwidth.

The simulated LTE downlink scenario adds a simulated channel and UE substituting the RF path of Fig. 4. It includes several VNFs: a data source, an eNodeB transmitter, the channel, a UE receiver, and a controller that allows selecting the working parameters, such as channel conditions and LTE link parameters.

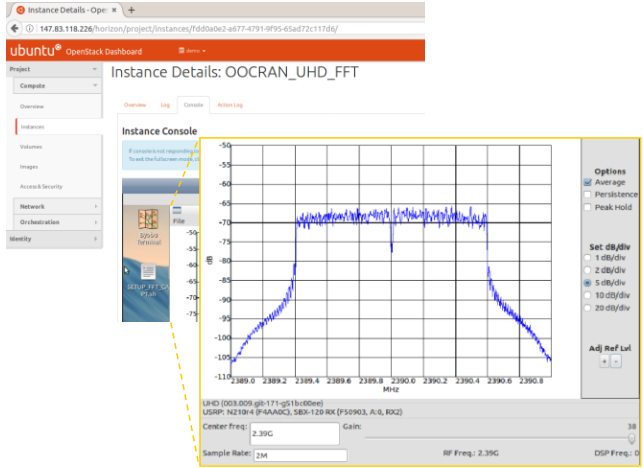

**Figure 5: Spectrum of the 1.4 MHz LTE downlink signal that is transmitted by the eNodeB instance and captured by another RRH.**

**Table 1: Time to setup a VWI**

| eNodeBs | Coverage Area | Time |
| --- | --- | --- |
| 1 | π*(30 m)² = 2826 m² | 30.12 s |
| 5 | 14,130 m² | 33.49 s |
| 10 | 28,260 m² | 45.87 s |
| 20 | 56,520 m² | 60.19 s |
| 30 | 84,780 m² | 84.63 s |

Once the VWI is configured and deployed for providing the specific service, it needs to be periodically adapted to the changing operational conditions, including changing traffic loads, channel impairments, and service requirements. One possibility is to create a tailored VWI for the new condition. The time that is required to set up and deploy a new VWI needs to be considered to ensure non-disruptive NS. All subscribers attached to the old VWI are released and remain disconnected from the network until the setup of the new infrastructure is completed.

The time it takes to set up and deploy a new VWI is a function of the number of eNodeBs needed to provide the desired coverage. Table 1 shows some figures. About a minute is needed to deploy a new VWI on the Barcelona Tech/EETAC campus with an area of about 58,241 m².

Fast swapping of a working VWI for another that better suits the new expected traffic load allows adapting the traffic capacity of the working VWI to the traffic demand. The relatively long times for VWI shutdown and redeployment calls for more sophisticated strategies to minimize the impact on user service perception. One solution is defining long periods for updates, e.g. one hour, or maintaining the old virtual network until the new network is established, introducing the notion of VWI soft handover. Another solution is creating a VWI repository. OOCRAN can then select the VWI that best matches the actual traffic demand, user distribution, and other conditions or requirements.

## 6 CONCLUSIONS

The discussion on how to apply ETSI NFV MANO to manage practical C-RAN deployments is still in its early stage. This paper has introduced the OOCRAN testbed and a methodology for setting up a wireless access network with real and simulated RF links. OOCRAN facilitates testing different infrastructure sharing methods and deployment strategies by providing monitoring and system analysis tools that do not jeopardize real-time execution. It enables creating and using a VWI repository for different types of experiments. The testbed is modular and can be easily extended with hardware or software to take into account additional environmental, service and network considerations, such as heterogeneous networks. It provides a platform for experimental research and education on virtualized wireless networks. Future work will address the creation of tailored and adaptive network services, analyze dynamic VWI deployment solutions and fundamental limitations, and tackle ultra-dense 5G networks, where efficient orchestration is critical.

## ACKNOWLEDGMENTS

This work has been partially supported by the Spanish Government, Ministerio de Ciencia e Innovación, through award number TEC2014-58341-C4-3-R and the NLnet Foundation.